\documentclass[submission,copyright,creativecommons]{eptcs}
\usepackage{graphics}
\usepackage{epsfig}
\usepackage{amsmath}
 \usepackage{setspace}
 \usepackage{graphicx}
\usepackage{subfig}
\usepackage{underscore}                                                                 

\begin{document}
\title{\bf{Comparison of Perfect and Quasi Werner States}}
\author{Fatima-Zahra Siyouri
\institute{ Laboratoire Equipe Sciences de la Mati\`ere et du Rayonnement}
\institute{Facult\'e des sciences, D\'epartement de Physique\\
Universit\'e Mohammed V - Agdal\\\
Rabat, Morocco}
\email{fatimazahra.siyouri@gmail.com}
\and
Fatima El Azzouzi
\institute{Facult\'e des sciences et techniques, D\'epartement de Math\'ematique\\
Universit\'e Sidi Mohamed Ben Abdellah\\\
Fes, Morocco}
\email{Fati.elazzouzi@gmail.com}
}
\def\titlerunning{Comparison of Perfect and Quasi Werner States}
\def\authorrunning{F-Z. Siyouri and F. El Azzouzi}
\maketitle

\begin{abstract}
 In this paper, we investigate comparatively the behaviors of quantum discord and concurrence for Werner states based on two bipartite entangled squeezed states. The maximally entangled squeezed states are regarded as a perfect-Werner states, while the non-maximally entangled squeezed states are called quasi-Werner states. We show that, the behavior of the quantum correlations keep unchanged for these two types of states. However, the quantum correlations amount presents in perfect-Werner states is very higher than that presents in quasi-Werner ones. Furthermore, we show that for large values of squeezed parameter the quasi-Werner states approaches the perfect-Werner states qualitatively and quantitatively.

\bigskip


\smallskip


\end{abstract}

\section{Introduction}

Quantum correlations have attracted a lot of interests recently, due principally to the general belief that they are important resources for various quantum tasks \cite{1,2,3,4,5,6,7,8,9,10,11,12,13}. Einstein et al \cite{14} and Schr\"odinger \cite{15} introduced the concept of quantum correlations called quantum entanglement \cite{16}, that can be considered for quantifying the quantum correlations present in various kinds of quantum systems \cite{17, 18,19,20}. Researchers have found that the entanglement is not the only type of quantum correlations, there are other quantum correlations even existing in separable mixed states. In 2001, Ollivier and Zurek  \cite{21,22} proposed quantum discord as a new type of the non-classical correlations \cite{16} defined as, the difference of the quantum versions of two classically equivalent expressions of the mutual information. The quantum discord is proposed to describe the quantum correlations, thus is not limited to entanglement.

	In this paper we investigate the quantum correlations behaviors. We consider systems that are described by Werner states based on bipartite entangled squeezed states and we work with the quantum discord as a measure of the total quantum correlations and with the concurrence as a measure of the special kind of quantum correlations that is entanglement. We show that, the behavior of the quantum correlations keep unchanged for these two types of states. However, the quantum correlations amount presents in perfect-Werner states is very higher than that presents in quasi-Werner ones. Furthermore, we show that for large values of squeezed parameter the quasi-Werner states approaches the perfect-Werner states and the amount of quantum correlation present in the quasi-Werner states becomes equal to those of perfect ones.

    The article is divided as follows. In section 2, a brief presentation of the quantum correlations is given. In section 3, we discuss comparatively the behavior of quantum discord and concurrence for quasi-Werner states and perfect-Werner states. In section 4, a conclusion of our results is given.

\section{Quantum correlations}

\subsection{ Quantum discord}

Let us consider a two random variables $X$ and $Y$. The classical mutual information between two discrete random variables in classical information theory is defined by:

\begin{equation}
\label{eq3}
 I(X:Y)=H(X)+ H(Y)- H(X,Y)\;,                                                                               
\end{equation}
where $ H(X) = -\sum\limits_{x \in  X} p(X) \log p(X)$  is the Shannon entropy \cite{23} associated to the random variable $X$ , and $H(X,Y) = - \sum\limits_{x \in  X , y \in  Y} p(X,Y) \log p(X,Y)$  is the joint entropy of $X$ and $Y$.

 From the joint probability $ p(X,Y)$ all probability distributions can be calculated as
 \begin{equation}
 \label{eq4}
 p(X)= \sum_{y \in  Y} p(X,Y=y) , \quad \hbox{and}\quad  p(Y)= \sum_{x \in  X} p(X=x,Y)\;,                                 
\end{equation}

 Another equivalent expression for classical mutual information is given by:
  \begin{equation}
  \label{eq5}
J(X:Y)=H(X)- H(X \mid Y)\;,                                                                                              
  \end{equation}   
where $ H(X \mid Y) $ is the conditional entropy \cite{23} that represents how much uncertainty we have on $X$ given the value of $Y$. It is given by:
\begin{equation}
\label{eq6}
\begin{split}
 H(X \mid Y) & = \sum_{y \in  Y} p(y) H(X\mid Y=y)  \\
             & = - \sum_{y \in Y} p(y) \sum_{x \in X} p(x \mid y) \log p(x \mid y)\;,                                      
 \end{split} 
\end{equation}
and it can be reduced to $ H(X \mid Y)= H(X ,Y)- H(Y) $.

However it was shown  \cite{21} that the quantum versions for these two, classically equivalent, expressions of mutual information are not equal. In fact the quantum version of equation  (\ref{eq3}) is straightforward and is obtained by replacing Shannon entropy $H(X)$ by the Von Neumann entropy $ S(\rho_X) $ \cite{29}:
 \begin{equation}
 \label{eq7}
 I(X:Y)= S(\rho_X) + S(\rho_Y) - S(\rho_{X,Y})                                 
\end{equation}

  Equation (\ref{eq5}) is less obvious to generalize \cite{21}, as it involves the conditional entropy on one subsystem given some information on the other one. Indeed, this can be achieved only by measuring the later, and the information gained about the subsystem depends on the measurement operators used.  In the case where one is restricted to performing projective measurements, the quantum analogue of the conditional entropy is given by:
  \begin{equation}
\label{eq8}
  S(\rho_{X \mid \left\{\prod\nolimits_j^Y\right\}}) = \sum_j  p_j S(\rho_{X \mid \prod\nolimits_j^Y})\;, 
 \end{equation}
 where $ \left\{\prod_{j}^{Y}\right\}$ is a complete set of orthonormal projection operators that act only on subsystem  $Y$,   $p_j= Tr_{X,Y} ({ \prod\nolimits_{j}^{Y}} \rho_{X,Y}) $ \quad \hbox{and} \quad $ \rho_{X \mid \prod\nolimits_{j}^{Y}} = \frac{1}{p_j} (\prod\nolimits_{j}^{Y}  \rho_{X,Y} \prod\nolimits_{j}^{Y})$.
 
 The most general local measurement operators
   $\left\{\prod_j^{Y}\right\}\equiv\left\{{|\Pi_1 \rangle \langle \Pi_1 |, |\Pi_2 \rangle \langle \Pi_2 |}\right\}$ can be easily constructed using the following quantum states: 
 $| \Pi_1 \rangle = \cos \theta| + \rangle + e^{i \Phi}\sin \theta |- \rangle$  \quad \hbox{and} \quad  $| \Pi_2 \rangle= \sin \theta |+  \rangle- e^{i \Phi}\cos \theta |- \rangle$.

With the above definitions in mind, the quantum analogue of equation (\ref{eq5}) becomes: 
\begin{equation}
 \label{eq9}                                                           
J(X:Y)_{\left\{\prod\nolimits_{j}^{Y}\right\}}= S(\rho_X)- S(\rho_{X \mid \left\{ \prod\nolimits_{j}^{Y}\right\}})\;
\end{equation}

In order to ensure that this last definition takes into account all classical correlations one must maximize it over all possible measurement basis $ {\left\{ \prod\nolimits_{j}^{Y}\right\}}$ of subsystem $Y$ \cite{16}.

Consequently, the quantum discord capturing all quantum correlations is defined as:
\begin{equation}
 \label{eq10}
\begin{split}
  D(X:Y)_{\left\{ \prod\nolimits_{j}^{Y}\right\}}& =I(X:Y)-\hbox{max}_{\left\{ \prod\nolimits_{j}^{Y}\right\}} \bigg[J(X:Y)_{\left\{\prod\nolimits_{j}^{Y}\right\}}\bigg]\\
                                                 & = S(\rho_Y) - S(\rho_{X,Y}) +\hbox{min}_{\left\{ \prod\nolimits_{j}^{Y}\right\}} S(\rho_{X \mid \left\{ \prod\nolimits_{j}^{Y}\right\}})\;,
\end{split} 
 \end{equation}
and the minimization now is carried over the variables $\theta$ and $\Phi$.

 It turns out that maximizing equation (\ref{eq8}) is the main difficulty in finding general analytic expressions for quantum discord present in arbitrary states. Indeed, exact analytical expressions are found only in a limited number of cases and the most general approach up to date was obtained for the so-called X-states \cite{24}. This approach has been successfully applied to understand and quantify the quantum correlations present in different types of systems\cite{30,31,32}.

\subsection{Concurrence}
 
The concurrence \cite{25} is one of the widely used measures of quantum entanglement in a bipartite state.

The Wootters concurrence of the density matrix $\rho_{XY}$ is defined as:

\begin{equation}
  \label{eq11} 
 C(\rho_{XY}) = \max  \left\{0 ,(\lambda_1 - \lambda_2 - \lambda_3 - \lambda_4)\right\}
 \end{equation}

where $\lambda_1 \geq \lambda_2 \geq \lambda_3 \geq \lambda_4$ are the square roots of the eigenvalues of the matrix $ \rho_{XY}\widetilde{\rho_{XY}}$.  The Spin Flipped density matrix $\widetilde{\rho_{XY}}$ is written as 

\begin{equation}
  \label{eq111} 
   \widetilde{\rho_{XY}}= ( \sigma_y \otimes  \sigma_y) \rho_{XY}^{\ast} ( \sigma_y \otimes \sigma_y)
 \end{equation}
With $\rho_{XY}^{\ast}$ is the complex conjugate of $\rho_{XY}$, and $\sigma_y$ is the Pauli matrix is  written  as
\begin{equation}
\label{eq12}
   \sigma_y=     
 \begin{pmatrix}
   0   &  -i \\
   i   &  0
\end{pmatrix}
\end{equation}

\section{ The  perfect and quasi Werner states}

Considering the squeezed state $ | \xi \rangle $, with parameter $r$, given by \cite{26} : 
 
 \begin{equation}
 \label{eq13}
| \xi \rangle = \frac{1}{\sqrt{\cosh r}}  \sum_0^{\infty} \frac{\sqrt{(2n)!}}{n!} (\frac{\tanh r}{2})^{n}
 |2n \rangle    
 \end{equation}

the two bipartite entangled squeezed state are written as :

 \begin{equation}
 \label{eq14}
  |\psi^{\pm} \rangle_{XY} = n_{\pm}[| \xi,\xi \rangle \pm |-\xi,-\xi \rangle]_{XY}\;,
 \end{equation}
  where $n_{\pm}$ is the normalization factor given as:
 \begin{equation}
 \label{eq15}
  n_{\pm}=[2 (1 \pm M^2)]^{-\frac{1}{2}}\;,
 \end{equation}
 with $ M = \frac{1}{\sqrt{\cosh 2 r}}$     

The states $ |\pm \xi \rangle $ are not orthogonal, a orthogonal basis is obtained by considering odd and even
coherent states $ | \pm \rangle $ given by:

 \begin{equation}
 \label{eq16}
  | \pm \rangle = N_{\pm}[| \xi \rangle \pm |-\xi \rangle]\;,
  \end{equation}
where $ N_{\pm}$ is given as
 \begin{equation}
 \label{eq17}
  N_{\pm} =[2(1 \pm M)]^{-\frac{1}{2}}.
\end{equation}

Considering the orthogonal basis of states $| \pm \rangle$, we express $|\psi^{\pm} \rangle_{XY} $ as follows: 
\begin{equation}
  \label{eq18}
 | \psi^{+} \rangle_{XY} = \frac{n^{+}}{2}[\frac{|+ , +\rangle}{N_{+}^{2} }+\frac{ |- ,- \rangle}{  N_{-}^{2}}]_{XY}\quad;\quad
 | \psi^{-} \rangle_{XY} = [\frac {|+ , -\rangle + |- , +\rangle}{\sqrt{2}}]\;,
  \end{equation}
  
Respectively, the Werner states \cite{27} based on the bipartite squeezed entangled states are written as:
\begin{equation}
 \label{eq19}
 \rho(\psi^{+},a)= (1-a) \frac{I}{4}+a \left| \psi^{+} \right\rangle \left\langle  \psi^{+} \right|\quad;\quad 
 \rho(\psi^{-},a)= (1-a) \frac{I}{4}+a \left| \psi^{-} \right\rangle \left\langle  \psi^{-} \right|\;,
\end{equation} 
 
$a$ plays the role of a mixing parameter, ranging from 0 to 1.
  
The states $ | \psi^{+} \rangle $ are non-maximally entangled, therefore, $\rho(\psi^{+},a)$ are known as quasi-Werner states. On the other hand, the states $ | \psi^{-} \rangle $ are maximally entangled states, therefore, $\rho(\psi^{-},a)$ are equivalent to perfect-Werner mixed states. The density matrices for the quasi-Werner states ($\rho(\psi^{+} ,a) $) and the perfect-Werner states are respectively written as:

\begin{equation}
 \label{eq20}
\rho_{XY}(\psi^{+},a)=      
  \begin{pmatrix}
 \frac{1-a}{4}+\bigg(\frac{a (1 + M)^{2}}{2(1 + M^{2})}\bigg) & 0 & 0 &    
  \bigg(\frac { a (1 + M) (1 - M)}{2(1 + M^{2})}\bigg) \\
    0 & \frac{1-a}{4} & 0 & 0 \\
   0 & 0 & \frac{1-a}{4} & 0 \\
   \bigg(\frac { a (1 + M) (1 - M)}{2(1 + M^{2})}\bigg) & 0 & 0 & \frac{1-a}{4}+\bigg(\frac{ a (1 - M)^{2}}{2(1 + M^{2})}\bigg)
\end{pmatrix}
 \end{equation}

	\begin{equation}
 \label{eq21}
\rho_{XY}(\psi^{-},a)=      
  \begin{pmatrix}
 \frac{1+a}{4}& 0 & 0 & \frac{a}{2} \\
    0 & \frac{1-a}{4} & 0 & 0 \\
   0 & 0 & \frac{1-a}{4} & 0 \\
  \frac{a}{2} & 0 & 0 & \frac{1+a}{4}
\end{pmatrix}
 \end{equation}

\section{Discussion}

In order to compare between quantum correlations for both quasi-Werner states and perfect-Werner states, we plotted their Variations with respect to the mixing parameter $a$ and the squeezed parameter $r$ in Fig. 1 and Fig. 2.

\begin{figure}[ht]
  \centering
  \subfloat[][$\rho(\psi^{+},a)$ with $r=0.7$]{\includegraphics[width=0.5\textwidth]{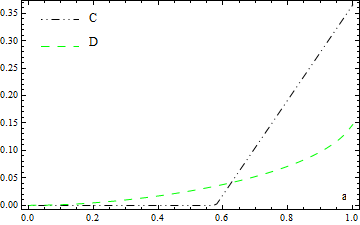}}
  \subfloat[][$\rho(\psi^{+},a)$ with $r=2.5$]{\includegraphics[width=0.5\textwidth]{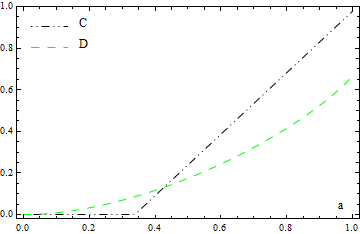}}\\
  \subfloat[][$\rho(\psi^{+},a)$ with $a=0.7$]{\includegraphics[width=0.5\textwidth]{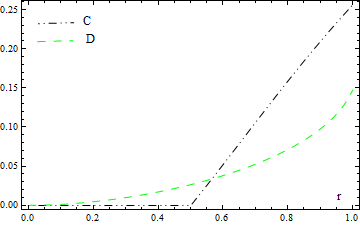}}
  \caption{ Variation of quantum discord and concurrence of quasi-Werner states with respect to the mixing parameter $a$ for fixed values of squeezed parameter $r$ (1-a and 1-b)and with respect to squeezed parameter $r$ for fixed values of $a$ (1-c) }\label{fig1}
\end{figure}

 Fig. 1-a and Fig. 1-c show that the quantum discord augments as $a$ and $r$ augment. So, the information is locally difficult to be accessible and cannot be obtained by distant independent observes without perturbing the system. Furthermore, it is seen that it attains its maximum value for the large values of $a$ and $r$ and it vanishes for very small values of these two parameters and therefore becomes equal to the classical correlations.  Moreover, it is shown  that the concurrence vanishes and then augments as the parameters $a$ and $r$ augment with higher rate than quantum discord starting from a given value of $a \geq 0.6$ and $r \geq 0.5$.  The quasi-Werner states are more separable whenever the parameters $r$ and $a$ are smaller. As a matter of fact, the concurrence gives us an idea about the degree of entanglement of the bipartite entangled squeezed states with respect to the parameter $r$ and the mixing parameter $a$. Increasing these parameters enhances the concurrence. Interestingly, we notice that entanglement is absent from the system for smal values of a ($ 0.15 \leq a \leq 0.6$) and r ($ 0.15 \leq r \leq 0.5$), while quantum discord is present. This indicates that quantum discord exits even when there's no entanglement. Which means that the discord is more robust than entanglement. Physically, this prove that the work in Ref. \cite{28}, which indicates the existence of quantum correlations other than entanglement, is valid only when the mixing parameter a has a small value and when we have continuous variables of smaller amplitudes for the used quantum states.
 
On the other hand, one can remark from  Fig. 1-a, Fig. 1-b and Fig. 1-c that the concurrence and the quantum discord curves are different, this is due to the fact that their definitions are fundamentally different. More precisely, the quantum discord given in (\ref{eq10}) has a nonlinear curve because of the minimization of the conditional entropy. \cite{21,22,16}.  While, the concurrence \cite{25} which is given as function of square root of eigenvalues of the matrix $ \rho_{XY}\widetilde{\rho_{XY}}$ (\ref{eq11}) such that, these eigenvalues are given as a set of scalars associated with linear system of equation for this reason the concurrence curve is linear.  It is worthwhile to notice that , in this study, the concurrence depends only on the parameters a and r, whereas, the quantum discord depends on the parameters a, r and the variables of minimisation $\theta$ and $\Phi$.

\begin{figure}[ht]
  \centering
  \subfloat[][$\rho(\psi^{-},a)$]{\includegraphics[width=0.5\textwidth]{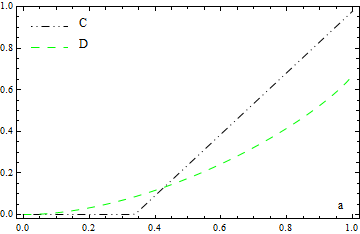}}
  \caption{ Variation of concurrence and quantum discord of perfect-Werner states with respect to the mixing parameter $a$  }\label{fig2}
\end{figure}

 Fig. 1-b shows that the quasi-Werner states contain more quantum discord which means that the quantum correlations are greater than the classical correlations and become maximally entangled for large values of squeezed parameter r ($r \geq 2.5$) and when the state is pure ($a =1$). This leads to the fact that in the presence of these two conditions the cryptography protocols using bipartite squeezed states will have good fidelity with a good probability of success.  which allows the detection of any eavesdropping measurement thanks to the perturbation. It is shown also from Fig. 2 that the behavior of both entanglement and quantum discord keep unchanged for the perfect-Werner states.  From these two figures, ( Fig.1-b and Fig.2) one can observe that for large values of the squeezed parameter $r$ the amount of both entanglement and quantum discord present in perfect-Werner states is equal to that present in quasi-Werner ones. Indeed, In this limit, we can conclude that for the large values of squeezed parameter $r$, the quasi-Werner states are almost equal to the perfect-Werner states qualitatively and quantitatively.

\section{Conclusion} 
 
 In this paper, we studied the quantum discord and quantum entanglement of Werner states based on two bipartite entangled squeezed states. It is revealed that, for both perfect-Werner states and quasi-Werner states, quantum discord and concurrence increase as the parameters $a$ and  $r$ are increased, but this increase in quantum correlation contributes more to quantum entanglement than to quantum discord. For quasi-Werner states $\rho_{XY}(\psi^{+},a)$, the calculations of quantum discord and concurrence depend mainly on the measurement basis as well as on the parameters $a$ and $r$, while for perfect-Werner states $\rho_{XY}(\psi^{-},a)$, it depend only on the parameter $a$. Both the quantum discord and concurrence are biggest for a large values of $r$ and $a$. So in our study, corresponding to our need in quantum communication and quantum computation, we can control the quantum correlations amount between two subsystems by controlling the measurement basis, the mixing parameter $a$ or the parameter $r$.

The non-zero discord guarantees the inaccessibility of information. On the other hand, quantum discord may exist even without entanglement. In fact the quantum information processing using the discord, it will be more performance than entanglement. The results obtained here using squeezed states also confirm the general results obtained in the literature. As a matter of fact, for maximally mixed states, no quantum correlations exist, so both entanglement and discord are null. Also, for pure Bell states, entanglement represent all the quantum correlations.This work allows us to describe the best characterization of Werner states Based on Bipartite Entangled Squeezed States, which gives more efficiency in quantum computation.

\bibliographystyle{eptcs}

\bibliography{generic}

\end{document}